\documentstyle[aps,twocolumn,epsf]{revtex}
\begin{document}
\title{Critical Dynamics of Contact Line Depinning}
\author{Deniz Erta\c s and Mehran Kardar}
\address{Department of Physics\\
Massachusetts Institute of Technology\\
Cambridge, Massachusetts 02139}
\date{\today}
\maketitle
\begin{abstract}
The depinning of a contact line is studied
as a dynamical critical phenomenon by a functional renormalization
group
technique.
In $D=2-\epsilon$ interface dimensions, the roughness exponent
is $\zeta=\epsilon/3$ to all orders in
perturbation theory. Thus, $\zeta=1/3$ for the contact
line, equal to the Imry-Ma estimate of Huse for the equilibrium
roughness. The dynamical exponent is
$z=1-2\epsilon/9+O(\epsilon^2)<1$,
resulting in unusual dynamical
properties. In particular, a characteristic distortion length
of the contact line depinning from a strong defect is predicted to
initially
increase faster than linearly in time.
Some experiments are suggested to probe
such dynamics.
\end{abstract}

Wetting phenomena and contact lines (CLs) appear in many
manufacturing processes that involve the spreading of a liquid on
a solid surface\cite{DeGennes}.
Some degree of control over the spreading of the liquid and the
corresponding creep of the CL is needed to optimize the
desired characteristics of such processes. In particular, it
is important to know the effect of surface roughness and
contaminants at microscopic to mesoscopic scales on CL dynamics.
Here, we study such effects at length scales
from $10^{-1}$ down to $10^{-7}$ centimeters. The upper length
scale is set by the droplet size or the capillary length
(due to gravitation), while the lower length scale is determined
by the characteristic size of the microscopic defects.

Surface impurities lead to {\it CL hysteresis}, i.e. a finite force
is
needed to start the fluid spreading.
Recently, the scaling exponents of a driven elastic interface subject
to
quenched impurities near a similar depinning treshold have been
calculated through a functional
renormalization-group (RG) treatment close to four interface
dimensions\cite{Nattermann,NF}. Here we apply
this method to calculate various scaling exponents for the
slowly advancing contact line. The distinction between the
two cases is that the CL is the termination of the liquid-vapor
interface.
We shall assume that the partially wetting fluid spreads sufficiently
slowly
on a heterogenous surface that the liquid-vapor interface
evolves adiabatically, i.e., the it responds
to changes in the CL shape instantenously.
In this case, fluctuations of the CL around
its time-averaged value reflect the competition between impurities
on the solid surface and the liquid-vapor surface tension.

\begin{figure}
\epsfxsize=2.9truein
\hskip 0.15truein\epsffile{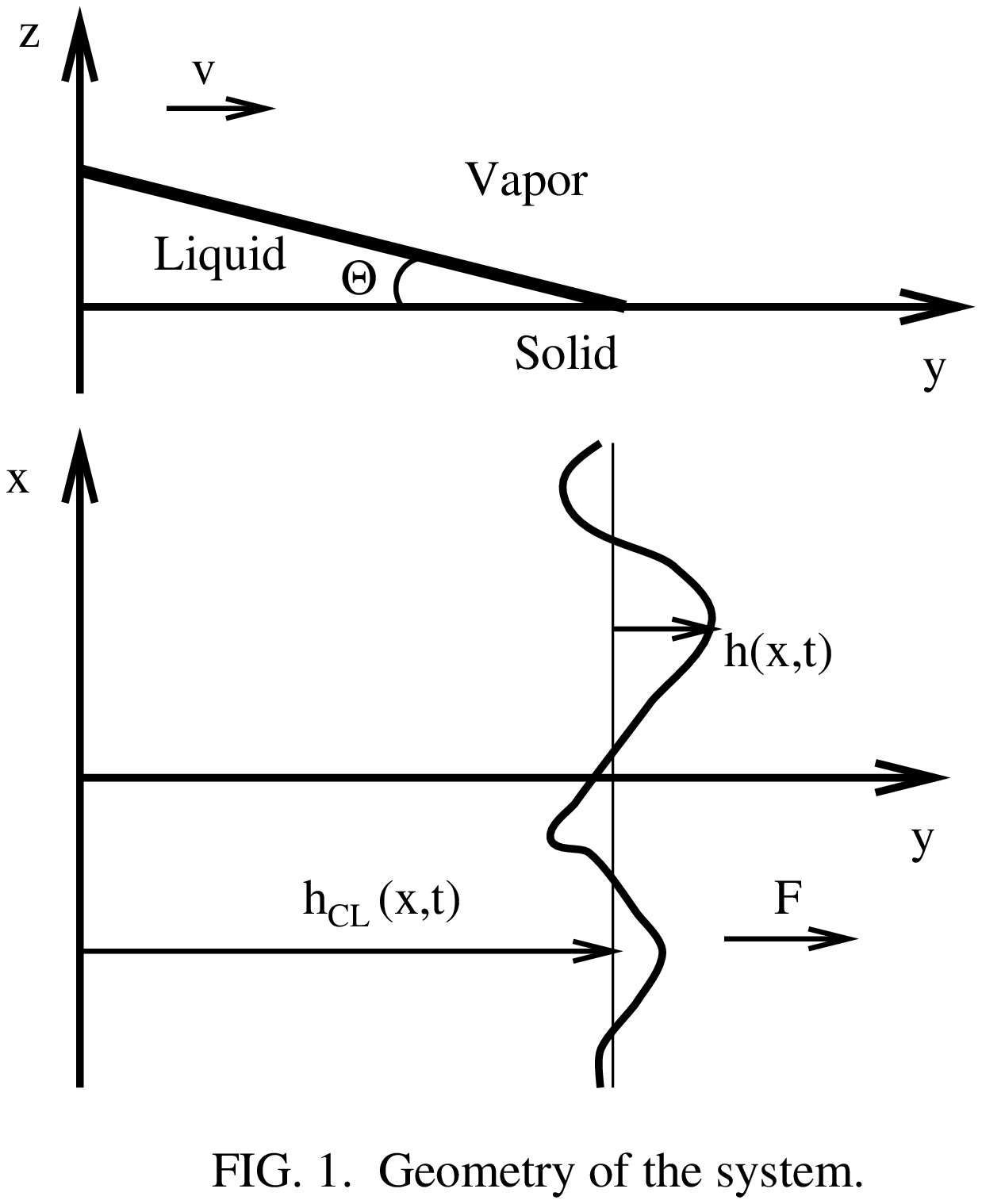}
\label{geometry}
\end{figure}

Consider a wetting front on a heterogenous surface in
the $x$-$y$ plane with the average orientation of the contact line in
the
$x$ direction as shown in Fig. \ref{geometry}.
In equilibrium on a {\it homogenous} interface, the
macroscopic contact angle $\Theta$
is determined by the Young condition,
\begin{equation}
\label{balance}
\gamma_{SV}-\gamma_{SL}-\gamma \cos\Theta = 0.
\end{equation}

In the above, $\gamma_{SV}$, $\gamma_{SL}$, and $\gamma$
are the interfacial tensions for the solid-vapor,
solid-liquid, and liquid-vapor interfaces, respectively.
The heterogeneities (aka defects) on the surface
can be modeled as fluctuations in the difference of local
interfacial energy densities,
\begin{equation}
f(x,y) =\gamma_{SV}(x,y)-\gamma_{SL}(x,y)
-\langle\gamma_{SV}-\gamma_{SL}\rangle.
\end{equation}
The average of $f$ is zero, while its correlations satisfy,
\begin{equation}
\langle f(x,y)f(x',y')\rangle  = \Delta(r/a),
\end{equation}
where $r^2=(x-x')^2+(y-y')^2$, $a$ is the characteristic
size of defects, and $\Delta$ is a function
that decays rapidly for large values of its argument.
(For a self-affinely rough surface, $\Delta$ may have a slow
algebraic decay\cite{DeGennes}. Such situations will not
be explored here.)
Fluctuations of the CL from its average position, $h_{CL}(t)=vt$,
are denoted by $h(x,t)$, and thus satisfy
\begin{equation}
\label{consistency}
\overline{\langle h(x,t)\rangle}=0.
\end{equation}

(The overline denotes a {\it time} average.)
The capillary energy associated with small
deformations of a CL was calculated by
Joanny and de Gennes\cite{Joanny}
(in the $\Theta\to0$ limit) as
\begin{eqnarray}
U_{\rm cap}&=& \frac{\gamma\Theta^2}{2}
\int_{\frac{2\pi}{a}<|q|<\frac{2\pi}{L}}
\frac{dq}{2\pi}|q||h(q)|^2 \nonumber \\
&=&\frac{\gamma\Theta^2}{2\pi}
\int_{a<|x-x'|<L} dxdx'\frac{h(x)h(x')}{(x-x')^2},
\end{eqnarray}
where $h(q)$ is the fourier transform of the contact line
profile at a given time, and $a$ and $L$ are the lower and upper
cutoff length scales mentioned
earlier. The unusual $q$-dependence of the
energy functional, and the resulting nonlocal dynamics,
reflects the fact that perturbations
of wavevector $q$ on the CL induce deformations
into the liquid-vapor interface up to a distance $|q|^{-1}$.

A random contribution to the liquid-surface energy comes from
the defects\cite{DeGennes},
\begin{equation}
U_{\rm rand}=\int_{-\infty}^{+\infty}dx\int_{-\infty}^{h_{CL}+h(x)}
dy\ f(x,y).
\end{equation}
If the heterogeneity is strong enough,
there is contact angle hysteresis\cite{DeGennes,Joanny}.
This arises from many metastable configurations of
the CL profile,
which are given by local minima of the free energy.
The surface tensions $\gamma_{SL},\gamma_{SV}$ in
Eq.(\ref{balance}) must then be interpreted as spatial
averages {\it over the CL position}. Thus, the macroscopic
contact angle will depend on the particular CL profile.
If  a macroscopic force of $F$ per unit length is exerted on the CL,
it will move with a finite velocity $v$ only for
$F>F_c$. This occurs when the metastable state with the largest
$\Theta$,  usually referred to as the advancing angle
$\Theta_a$\cite{DeGennes,Joanny}, becomes unstable.
For small capillary numbers, $\Theta\approx\Theta_a$.

As the CL slowly advances, energy is dissipated through various
mechanisms\cite{DeGennes}. If the dissipation in the vicinity
of the CL dominates for small $v$, the drag force at a point $x$
on the CL is simply related to the local velocity
$v+\partial_th(x,t)$
through a microscopic mobility $\mu$. In this case, the equation
of motion for the CL is obtained by equating the drag force to
the applied force, $-\delta U/\delta h(x)$, as
\begin{eqnarray}
\label{motion}
\mu^{-1}\left(v+\frac{\partial h(x,t)}{\partial t}\right) &=&
-\frac{\gamma\Theta^2}{\pi}\int_{a<|x-x'|<L}
dx'\frac{h(x',t)}{(x-x')^2} \nonumber \\
& &\ +f[x,vt+h(x,t)]+F.
\end{eqnarray}
In the above formula, $v$ is determined self-consistenly by
enforcing Eq. (\ref{consistency}).
Note that Eq.(\ref{motion}) is only valid for $v>0$, and
$F\to F_c$ from above when $v \to 0$. The dynamical contact
angle $\Theta$ is given by the force balance equation,
\begin{equation}
F-F_c=\gamma(\cos\Theta_a-\cos\Theta).
\end{equation}
A recent experiment on depinning from a single defect
\cite{Marsh} is in agreement with the deterministic $(f=0)$
form of this dynamics: Upon depinning from the defect,
the time evolution of the CL profile is given by
\begin{equation}
h(x,t)\sim\ln\left(\frac{x^2+(ct)^2}{L^2}\right),
\end{equation}
where the width of the logarithmic profile increases
linearly  with a characteristic
velocity $c=\gamma\Theta^2\mu$ after depinning.

By analogy to similar previously studied systems
\cite{Nattermann,NF,CDW}, for $F$ slightly above treshold,
we expect the average velocity of the contact line to scale as
\begin{equation}
\label{velocity}
v=A(F-F_c)^\beta,
\end{equation}
where $\beta$ is the {\it velocity exponent}, and $A$ is a
nonuniversal constant.
Superposed on the steady advance of the CL
are rapid ``jumps" as portions of the line
depin from strong pinning centers. Such jumps are similar
to {\it avalanches} in other slowly forced systems, and have a
power-law distribution in size, cut-off at a characteristic
correlation length $\xi$. On approaching the threshold,
$\xi$ diverges as
\begin{equation}
\xi\sim (F-F_c)^{-\nu},
\end{equation}
defining a {\it correlation length exponent} $\nu$. At length scales
smaller than $\xi$, the interface is self-affine, i.e.
with correlations satisfying the dynamic scaling form
\begin{equation}
\langle\left(h(x,t)-h(x',t')\right)^2\rangle=(x-x')^{2\zeta}g\left(
{{t-t'}\over{(x-x')^z}}\right),
\end{equation}
where $\zeta$ and $z$ are the {\it roughness} and {\it dynamic}
exponents, respectively. The scaling function $g$
goes to a constant as its argument approaches 0, $\zeta$ is
the wandering exponent of an instantenous CL profile, and
$z$ relates the average lifetime of an avalanche to its size
by $\tau(\xi)\sim\xi^z$.

In many aspects, Eq.(\ref{motion})
is similar to the model analyzed by Narayan and Fisher
using the formalism of Martin, Siggia, and Rose\cite{MSR} (MSR),
through an expansion around a mean field solution\cite{NF}.
To use this method, it is necessary to generalize
to a $D$-dimensional interface for
a systematic computation of the exponents through an $\epsilon$
expansion. Introducing an auxiliary field $\hat h(x,t)$, the
MSR generating functional is,
\begin{eqnarray}
\label{MSRZ}
Z&=&\int [dh][d\hat h]  \exp \left\{ i\int{d^D{\bf x}\,dt\,\hat
h({\bf x},t){\cal F}[h,f]}\right\}, \\
&{\cal F}&[h,f]=\mu^{-1}\left(\partial_t h({\bf x},t)+v\right)
-f[{\bf x},vt+h({\bf x},t)]-F \nonumber \\
& & \quad+{\gamma\Theta^2}\int{d^D{\bf x'}K_D({\bf x}-{\bf x'})
h({\bf x'},t)}. \nonumber
\end{eqnarray}
In the above, $K_D({\bf x})\;(\sim|{\bf x}|^{-(D+1)}$ for $a<|{\bf x}|<L$)
is the generalized interaction kernel in $D$ dimensions.
Mean field (MF) theory is obtained by replacing the capillary forces
on a portion of the CL with Hookean springs. This gives
\begin{equation}
\mu^{-1}\left(\frac{dh_{MF}}{dt}+v\right)=-kh_{MF}(t)+f[vt+h_{MF}(t)]+F,
\end{equation}
independently for each ${\bf x}$, and identical
to the mean field model analyzed by NF\cite{NF,CDW}.
Here again, $v(F)$ is determined by the self-consistency
condition $\overline{\langle h_{MF}\rangle}=0$.

The MF solution depends on the type of irregularity: For
smoothly varying $f(x,y)$, $\beta_{MF}=3/2$, whereas for ``mesa"
defects, i.e. cusped $f(x,y)$, $\beta_{MF}=1$\cite{CDW,Robbins}.
Some recent experiments on reasonably clean surfaces by Str\"om
et al.\cite{Strom} are in agreement with Eq.(\ref{velocity}) for
$\beta=3/2$ in the low capillary number regime. Even though
it is generally believed that the scaling relation in
Eq.(\ref{velocity})
holds\cite{Seebergh}, it is clear that the prefactor is
nonuniversal, and various experiments to date have obtained
widely varying $v$-$F$ behaviors\cite{Hayes}. To our knowledge,
there are no experiments so far that have systematically investigated
strongly heterogenous surfaces with reasonable accuracy.

Following the treatment of NF\cite{NF,CDW},
we use the mean field solution for cusped potentials,
anticipating jumps with velocity of $O(1)$, in which case
$\beta_{MF}=1$. After rescaling and averaging over impurity configurations,
we arrive at a generating functional whose low-frequency form is
\begin{eqnarray}
\overline Z&=& \int[dH][d\hat H]\exp(-S), \nonumber \\
\label{S}
S&=&\int d^D{\bf x}\,dt\, \left[F-F_{MF}(v)\right] \hat H({\bf x},t)
\nonumber
\\
& & + \int {d\omega \over 2\pi}
{d^D{\bf q} \over (2\pi)^D} \hat H({-\bf q},-\omega)(-i\omega+|{\bf
q}|)H({\bf
q},\omega)
\nonumber \\
& & -{1 \over 2}\int d^D{\bf x}\,dt\,dt'\, \hat H({\bf x},t)\hat
H({\bf x},t')
\nonumber \\
& & \qquad\quad \times C\left[vt-vt'+H({\bf x},t)-H({\bf
x},t')\right].
\end{eqnarray}
In the above expressions, $H$ and $\hat H$ are coarse-grained
forms of $h$ and $\hat h$, in the sense that response
and correlation functions of $h$, $\hat h$ are the same as those
of $H$, $\hat H$ at low frequencies. $F$ is adjusted to satisfy
the condition $\langle H\rangle=0$. The function $C(v\tau)$ is
initially the mean-field correlation function
$\langle h_{MF}(t)h_{MF}(t+\tau)\rangle$.

All non-gaussian terms in the generating
functional $\overline Z$, including ones that are not shown
in Eq.(\ref{S}), decay away at large length and time scales
for $D>D_c=2$. Thus, for $D>D_c$, exponents can be easily
determined from
the gaussian theory as $z_0=\nu_0=\beta_0=1$. The interface
is flat at long length scales $(\zeta_0=(2-D)/2)$.

At $D=D_c$, there are an infinite number of marginal operators
that can be cast into a single marginal function
for $v\to 0$. The RG is carried out by integrating over
a momentum shell and all frequencies, followed by a scale
transformation $x\to bx$, $t\to b^zt$, $H\to b^\zeta H$, and
$\hat H\to b^{\theta-D} \hat H$, where $b=e^l$.
For $D<D_c$, there are a number of exact exponent
relations\cite{NF}:
\begin{eqnarray}
\beta&=&(z-\zeta)\nu,\\
\label{exp1}
\nu&=&1/(z+\theta),\\
\label{exp2}
\nu&=&1/(1-\zeta).
\end{eqnarray}
The first result comes from the scaling of nonlinear
response functions, requiring $H$ to scale as $vt$. The second
relation follows from the renormalization of $F-F_{MF}(v)$
through the first
term in Eq.(\ref{S}). Finally, the last relation results from
an exact statistical symmetry of the system that fixes
the form of the static ($\omega =0$) response function
$\langle \partial h(x')/\partial f(x)\rangle$.
These relations determine all exponents in terms of
$\zeta$ and $z$. The diagrams appearing in the renormalization
of $C(u)$ are exactly the same as in Ref.\cite{NF}, with
the exception of the form of the bare propagator, which is
$(-i\omega+|q|)^{-1}$ instead of $(-i\omega+q^2)^{-1}$.
Thus, the renormalization of $C$ in $D=2-\epsilon$ interface
dimensions, computed to one-loop order,
gives the recursion relation,
\begin{eqnarray}
\label{RR}
{\partial C(u) \over \partial l}=[\epsilon &+& 2\theta
+2(z-1)]C(u)+\zeta u\frac{dC(u)}{du} \nonumber \\
 &-& {1\over 2\pi}\frac{d}{du}\left\{\left[C(u)-C(0)\right]
\frac{dC(u)}{du}\right\}.
\end{eqnarray}
NF showed that all higher order diagrams contribute to the
renormalization of $C$ as
total derivatives with respect to $u$, thus, integrating
Eq.(\ref{RR})
at the fixed-point solution $\partial C^*/\partial l = 0$,
together with Eqs.(\ref{exp1}) and (\ref{exp2}),
gives $\zeta=\epsilon/3$ to all orders in $\epsilon$,
provided that $\int C^*\neq 0$. For the case of a CL $(D=1)$,
we thus obtain
\begin{equation}
\zeta=1/3,\quad{\rm and}\quad \nu=3/2.
\end{equation}
This value of the roughness exponent coincides with the Imry-Ma
estimate of Huse given in Ref.\cite{DeGennes}
 for the {\it equilibrium} roughness. This is a consequence
of the fact that $C(u)$ remains short-ranged
upon renormalization, implying the absence of anomalous
contributions to $\zeta$.

To calculate the dynamical exponent $z$, we need
the fixed-point function $C^*$, which is obtained
only to $O(\epsilon)$ from the above analysis. Furthermore,
the calculation requires some knowledge of the high-frequency form
of response and correlation functions, i.e., a low-frequency
analysis of Eq.(\ref{S}) is insufficient to describe
scaling properties of the
system. Nevertheless, NF calculate the exponent
$\theta$ to $O(\epsilon)$ as $\theta+\zeta=(\epsilon-\zeta)/3
+O(\epsilon^2)$. The computation for the CL gives the same result,
yielding
\begin{eqnarray}
z&=&1-2\epsilon/9+O(\epsilon^2), \\
\beta&=&1-2\epsilon/9+O(\epsilon^2).
\end{eqnarray}
Note that $\beta<z$ even though the difference is $O(\epsilon^2)$.
Nattermann et al\cite{Nattermann} obtain the same results
to $O(\epsilon)$ by directly averaging the MSR generating
function in Eq.(\ref{MSRZ}).
The treatment of NF has the advantage of expanding
around a solution with $F_c\neq 0$, resulting in a better behaved
theory. In particular, systems that are described by vector fields
instead of scalar fields, like the treshold critical dynamics of
flux lines in three
dimensions, require the use of their approach, since the expansion
has to be made around a solution with $F_c \neq 0 $\cite{EKDepin}.

\begin{figure}
\epsfxsize=2.9truein
\hskip 0.15truein\epsffile{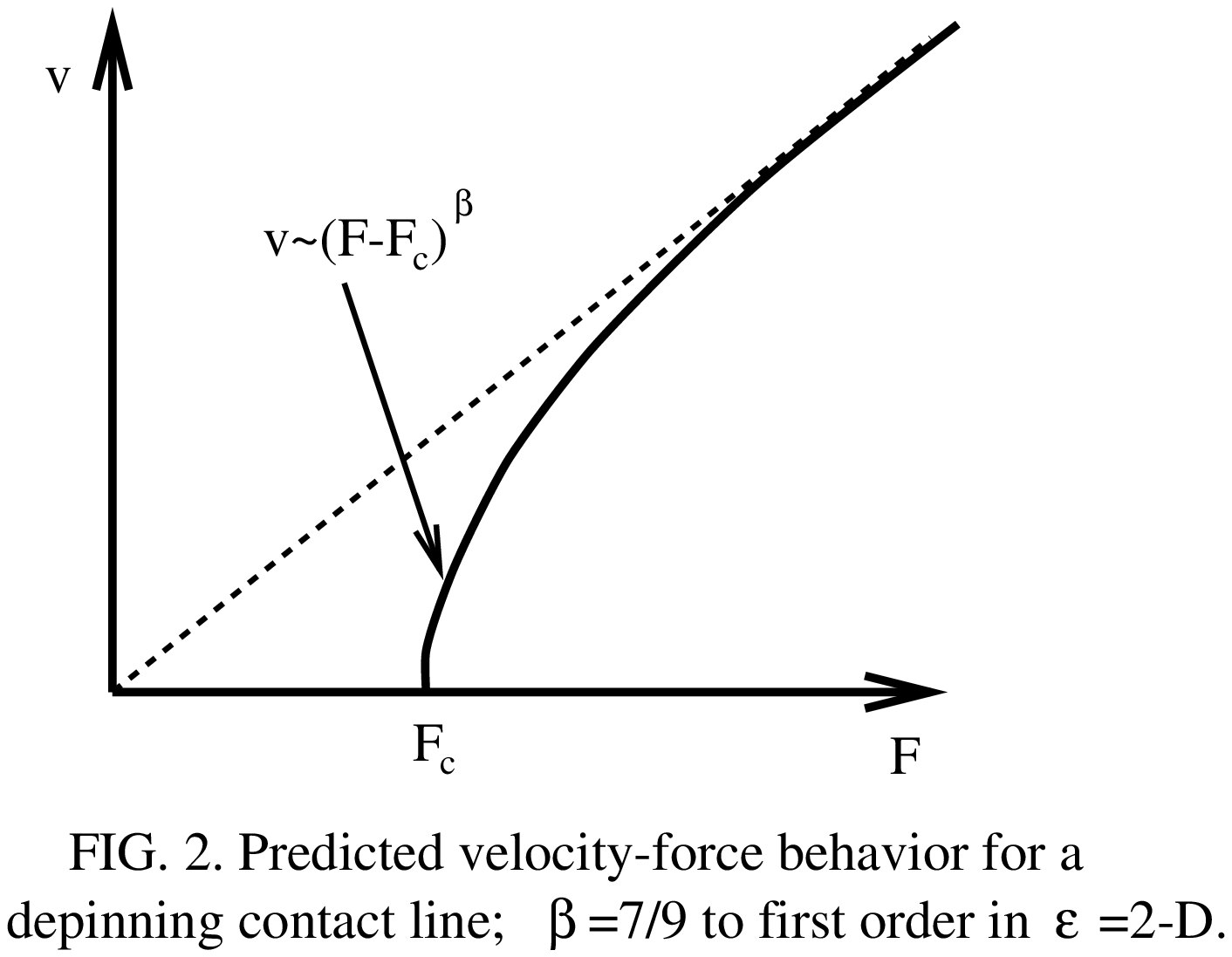}
\label{VFgraph}
\end{figure}

The above scaling exponents describe a CL
advancing at low capillary number on a dirty
surface, and also apply to surfaces
with microscopic roughness, i.e. with short range
slope-slope correlations.
The roughness exponent of  $\zeta=1/3$, equal
to its equilibrium value, can
be directly measured by microscopic imaging of a slowly advancing
CL. A number of macroscopic experimental tests are also
possible: the velocity of the CL is given by
by $v\sim(F-F_c)^\beta$ with $\beta<1$ (See Fig. \ref{VFgraph}).
The root mean square width of the CL profile should increase
like $W\sim\xi^\zeta\sim v^{-\zeta/\nu\beta}$ as the threshold is
approached. These relations can be checked through
tensiometric measurements\cite{Hayes}, where the capillary
force on a solid immersed into a liquid is measured directly.

A dynamical exponent of $z<1$ suggests that the relaxation of the
CL is very different on a dirty surface. Upon depinning from a
strong defect, the width of the logarithmic CL profile
initially grows faster than linearly in time, in other words,
with a characteristic velocity that increases with time:
$c(t)\sim t^{(1-z)/z}$.
This is, of course, not physical at arbitrarily large length
and time scales. We have assumed that the liquid-vapor interface
retains its equilibrium shape determined by the CL profile
at all times. This picture will break down as characteristic
velocities become comparable to capillary wave velocity
$c_{\rm cap}=\sqrt{\gamma/\rho}$ of the liquid-vapor interface.
The scaling regime in between should still be accessible to
 experiments in which the depinning from a single strong
defect is observed on a dirty or rough surface.

Finally, NF carried their analysis further to speculate exponents
{\it below} the treshold when the driving force is increased
monotonically towards $F_c$\cite{NF}. In particular, they postulated
an avalanche distribution
\begin{equation}
{\rm Prob(width\ of\ avalanche} > \ell) \approx \frac{1}{\ell^\kappa}
\hat\rho(\ell/\xi_-),
\end{equation}
with $\xi_-\sim(F_c-F)^{\nu_-}$, and the mean polarizability
\begin{equation}
\chi_\uparrow=\frac{d\langle \int h\rangle}{dF_\uparrow}
\sim(F_c-F)^{-\gamma}.
\end{equation}
If $\nu_-=\nu$, and the scaling hypothesis holds for the CL,
it then follows that $\kappa=D-1/\nu_-$,
and $\gamma=1+\zeta\nu_-$. These exponents are then given exactly
by $\gamma=\nu_-=3/2$, and $\kappa=1/3$ in $D=1$.

We thank O.~Narayan for various discussions and communicating
his results. This research was supported by grants from the NSF
(DMR-93-03667 and PYI/DMR-89-58061), and the MIT/INTEVEP
collaborative program.

\end{document}